\documentclass[12pt,preprint]{aastex}

\newcommand{\Fermic}{\textit{Fermi}}
\newcommand{\Fermi}{\Fermic\ }
\newcommand{\FermiLATc}{\Fermic\ LAT}
\newcommand{\FermiLAT}{\FermiLATc\ }
\newcommand{\FermiGSTc}{\textit{Fermi Gamma-ray Space Telescope}} 
\newcommand{\FermiGST}{\FermiGSTc\ }
\newcommand{\Suzakuc}{\textit{Suzaku}}
\newcommand{\Suzaku}{\Suzakuc\ }
\newcommand{\Swiftc}{\textit{Swift}}
\newcommand{\Swift}{\Swiftc\ }
\newcommand{\SwiftXRTc}{\textit{Swift-XRT}}
\newcommand{\SwiftXRT}{\SwiftXRTc\ }
\newcommand{\UVOTc}{\textit{UVOT}}
\newcommand{\UVOT}{\UVOTc\ }
\newcommand{\XRTc}{\textit{XRT}}
\newcommand{\XRT}{\XRTc\ }
\newcommand{\RXTEc}{\textit{RXTE}}
\newcommand{\RXTE}{\RXTEc\ }
\newcommand{\Einsteinc}{\textit{Einstein}}

\newcommand{\ROSATc}{\textit{ROSAT}}

\newcommand{\ASCAc}{\textit{ASCA}}
\newcommand{\ASCA}{\ASCAc\ }
\newcommand{\BeppoSAXc}{\textit{BeppoSAX}}
\newcommand{\BeppoSAX}{\BeppoSAXc\ }
\newcommand{\XMMNewtonc}{\textit{XMM-Newton}}
\newcommand{\XMMNewton}{\XMMNewtonc\ }
\newcommand{\radioc}{\textit{radio}}

\newcommand{\opticalUVc}{\textit{optical-UV}}

\newcommand{\Xrayc}{\textit{X-ray}}

\newcommand{\VHEgammarayc}{\textit{VHE gamma ray}}

\shorttitle{PG\,1553+113}
\shortauthors{{\it{Fermi}}\/LAT Collaboration}

\begin{document}
\bibliographystyle{apj}

\title{\Fermi Observations of the Very Hard Gamma-ray Blazar PG\,1553+113}

\author{
A.~A.~Abdo\altaffilmark{1,2}, 
M.~Ackermann\altaffilmark{3}, 
M.~Ajello\altaffilmark{3}, 
W.~B.~Atwood\altaffilmark{4}, 
M.~Axelsson\altaffilmark{5,6}, 
L.~Baldini\altaffilmark{7}, 
J.~Ballet\altaffilmark{8}, 
G.~Barbiellini\altaffilmark{9,10}, 
D.~Bastieri\altaffilmark{11,12}, 
K.~Bechtol\altaffilmark{3}
% ...
}
\altaffiltext{1}{Space Science Division, Naval Research Laboratory, Washington, DC 20375, USA}
\altaffiltext{2}{National Research Council Research Associate, National Academy of Sciences, Washington, DC 20001, USA}
\altaffiltext{3}{W. W. Hansen Experimental Physics Laboratory, Kavli Institute for Particle Astrophysics and Cosmology, Department of Physics and SLAC National Accelerator Laboratory, Stanford University, Stanford, CA 94305, USA}
\altaffiltext{4}{Santa Cruz Institute for Particle Physics, Department of Physics and Department of Astronomy and Astrophysics, University of California at Santa Cruz, Santa Cruz, CA 95064, USA}
\altaffiltext{5}{Department of Astronomy, Stockholm University, SE-106 91 Stockholm, Sweden}
\altaffiltext{6}{The Oskar Klein Centre for Cosmoparticle Physics, AlbaNova, SE-106 91 Stockholm, Sweden}
\altaffiltext{7}{Istituto Nazionale di Fisica Nucleare, Sezione di Pisa, I-56127 Pisa, Italy}
\altaffiltext{8}{Laboratoire AIM, CEA-IRFU/CNRS/Universit\'e Paris Diderot, Service d'Astrophysique, CEA Saclay, 91191 Gif sur Yvette, France}
\altaffiltext{9}{Istituto Nazionale di Fisica Nucleare, Sezione di Trieste, I-34127 Trieste, Italy}
\altaffiltext{10}{Dipartimento di Fisica, Universit\`a di Trieste, I-34127 Trieste, Italy}
\altaffiltext{11}{Istituto Nazionale di Fisica Nucleare, Sezione di Padova, I-35131 Padova, Italy}
\altaffiltext{12}{Dipartimento di Fisica ``G. Galilei'', Universit\`a di Padova, I-35131 Padova, Italy}
% ...

\begin{abstract}

We report the observations of PG\,1553+113 during the first
$\sim$200\,days of \FermiGST science operations, from 4 August 2008 to
22 February 2009 (MJD 54682.7-54884.2). This is the first detailed
study of PG\,1553+113 in the GeV gamma-ray regime and it allows us to
fill a gap of three decades in energy in its spectral energy
distribution. We find PG\,1553+113 to be a steady source with a hard
spectrum that is best fit by a simple power-law in the \Fermi energy
band. We combine the \Fermi data with archival radio, optical, X-ray
and very high energy (VHE) gamma-ray data to model its broadband
spectral energy distribution and find that a simple, one-zone
synchrotron self-Compton model provides a reasonable fit. PG\,1553+113
has the softest VHE spectrum of all sources detected in that regime
and, out of those with significant detections across the \Fermi energy
bandpass so far, the hardest spectrum in that energy regime. Thus, it
has the largest spectral break of any gamma-ray source studied to
date, which could be due to the absorption of the intrinsic gamma-ray
spectrum by the extragalactic background light (EBL). Assuming this to
be the case, we selected a model with a low level of EBL and used it
to absorb the power-law spectrum from PG\,1553+113 measured with
\Fermi (200\,MeV\,-\,157\,GeV) to find the redshift which gave the
best fit to the measured VHE data (90\,GeV\,-\, 1.1\,TeV) for this
parameterisation of the EBL. We show that this redshift can be
considered an upper limit on the distance to PG\,1553+113.

\end{abstract}

\keywords{Gamma rays: observations --- BL Lacertae objects: individual
  (PG\,1553+113)}

\maketitle

\section{Introduction}
\label{SEC:INTRO}

PG\,1553+113 is a high-frequency peaked BL Lacertae object (HBL;
\citealp{Falomo:1994:SpecBlazars};
\citealp{Beckmann:2002:BeppoSAXBLLacs}) whose redshift remains unknown
despite continued efforts. Like all BL Lacs, we find its spectral
energy distribution (SED) to have a double-peaked shape (in
${\nu}F{\nu}$ representation) that can thus be parameterised with four
characteristic slopes. It has been detected from radio through hard
X-rays and also in the very high energy (VHE;
$E$\,$\gtrsim$\,100\,GeV) regime up to energies above 1\,TeV
(\citealp{HESS:2006:PG1553Detection};
\citealp{MAGIC:2007:PG1553Detection}). With these data three of the
four components of its SED were sampled, namely, the rising and
falling edges of the low-energy peak ($\sim$10$^{-6}$\,eV - 30\,keV)
and the falling edge of the high-energy peak ($\sim$90\,GeV -
1\,TeV). We report here the first detailed analysis of the rising
high-energy portion and, crucially, the high-energy peak, of the
PG\,1553+113 SED. These data, from observations made by the \FermiGST
Large Area Telescope (\FermiLATc; \citealp{Atwood:2009:LAT}) during
its first $\sim$200 days of operation, are combined with data from
other wavebands to construct and model the broadband SED of
PG\,1553+113 in Section~\ref{SEC:MODELING}.

Discovered and classified as a BL Lacertae object (BL Lac) by
\citet{Green:1986:PGCatalog}, PG\,1553+113 is located at R.A. of
$\alpha_{J2000}$ = 15h55m43.04s and declination of $\delta_{J2000}$ =
+11d11m24.4s in the constellation of Serpens Caput. The logarithmic
ratio of its 5\,GHz radio flux, $F_{5GHz}$, to its 2\,keV X-ray flux,
$F_{2keV}$, has been found to range from log($F_{2keV}$/$F_{5GHz}$) =
$-4.99$ to $-3.88$ (\citealp{Osterman:2006:PG1553mwl};
\citealp{Rector:2003:RadioHBL}). The high value of this ratio places
PG\,1553+113 at times among the most extreme of the HBLs; a BL Lac is
classified as extreme when it has log($F_{2keV}$/$F_{5GHz}$) $\geq$
-4.5 \citep{Rector:2003:RadioHBL}. A number of the TeV blazars have,
at one time or another, exhibited fluxes that place them in this
extreme category (e.g., 1ES\,0229+200, H\,1426+428, 1ES\,1959+650;
\citealp{Rector:2003:RadioHBL}).

In the radio band, PG\,1553+113 has been detected at different mean
flux levels. Its 4.8\,GHz flux, for example, has ranged from 180 to
675\,mJy (\citealp{Bennett:1986:PG1553SEDned};
\citealp{Gregory:1991:87GB}; \citealp{Becker:1991:GHzSources};
\citealp{Perlman:2005:XraySpectra};
\citealp{Osterman:2006:PG1553mwl}). Its flux between 4.8 and 14.5\,GHz
was found to be variable on month timescales during the observations
of \citet{Perlman:2005:XraySpectra} and
\citet{Osterman:2006:PG1553mwl}. VLBA observations have resolved a jet
extending at least 20\,pc to the northeast of PG\,1553+113
\citep{Rector:2003:RadioHBL}. No evidence for superluminal motion has
been reported in the literature to date; multi-epoch VLBA monitoring
has commenced
recently\footnote{http://web.whittier.edu/gpiner/research/index.htm}.

PG\,1553+113 is a bright optical source with $V$-band magnitude of
$V_o\,\sim\,14$ (\citealp{Falomo:1990:PG1553};
\citealp{Osterman:2006:PG1553mwl}). Observations taken between 1986
and 1991 with the ESO telescopes found its spectral index, $\alpha_o$,
to remain almost constant ($\alpha_o\,\sim\,-1$) and its magnitude to
vary by ${\Delta}V_o\,=\,1.4$ \citep{Falomo:1994:SpecBlazars}. Low
levels of optical variability were seen by
\citet{Osterman:2006:PG1553mwl} during their 2003 observing
campaign. Limits on the magnitude of its host galaxy will be discussed
later.

PG\,1553+113 is also a bright X-ray source that has been observed by
most X-ray missions (\Einsteinc, \ROSATc, \ASCAc, \BeppoSAXc, \RXTEc,
\XMMNewtonc, \Swift and \Suzakuc). Although it has been detected at a
number of different flux levels by these observatories, no evidence
for strong or fast (sub-hour) flux variability has been observed at
X-ray energies \citep{Reimer:2008:PG1553Suzaku}. The \Suzaku
observations performed in July 2006 \citep{Reimer:2008:PG1553Suzaku}
provide the highest energy X-ray measurement so far obtained for
PG\,1553+113 at $\sim 30$\,keV with a 10-30\,keV flux of $1.35$ x
$10^{-11}$\,erg\,cm$^{-2}$\,s$^{-1}$. No evidence for spectral
hardening up to these energies was found in these data, indicating
that all of the X-ray emission detected was due to synchrotron
emission. The 2-10\,keV fluxes measured by different X-ray missions
are shown in Table~\ref{TAB:X-RAY}, the highest being $6.9$ $\times$
$10^{-11}$\,erg\,cm$^{-2}$\,s$^{-1}$ in October 2006 (\SwiftXRTc;
\citealp{Tramacere:2007:SwiftTeV}). The X-ray flux was found to double
over a period of 10 days during the 3-week RXTE observing campaign of
\citet{Osterman:2006:PG1553mwl}. Despite the variations of a factor of
approximately 20 in the 2-10\,keV X-ray flux, the measured spectral
properties of PG\,1553+113 at these energies, also listed in
Table~\ref{TAB:X-RAY}, have not changed significantly over the course
of the X-ray observations. It exhibits spectral curvature that can be
well-described with either a broken power law or a log-parabolic
shape. In the \Suzaku data this curvature extends into the hard X-ray
band ($\leq 30$\,keV) and steepening of the spectrum above
$\sim10$\,keV beyond that predicted by either the broken power-law or
log-parabolic model is observed \citep{Reimer:2008:PG1553Suzaku}.

In the high energy (HE; $E$ $\sim$ 100\,MeV - 100\,GeV) gamma-ray
regime, PG\,1553+113 was not detected by EGRET. An upper limit of
$F{_{EGRET}}$ $<$ 9.97 $\times$ 10$^{-8}$\,cm$^{-2}$\,s$^{-1}$ above
100\,MeV was derived based on the summed exposure of cycles 1, 2, 3
and 4 \citep{Hartman:1999:EGRET3EG}. At higher energies, in the VHE
regime, PG\,1553+113 is a confirmed gamma-ray emitter. First detected
by H.E.S.S. \citep{HESS:2006:PG1553Detection} and subsequently
confirmed by MAGIC \citep{MAGIC:2007:PG1553Detection}, PG\,1553+113
has a flux that is approximately 3\% that of the Crab Nebula at these
energies. The combined 2005-2006 H.E.S.S. data follow a power law with
photon index of $\Gamma_{VHE}$=4.46\,$\pm$\,0.34 between
$\sim$225\,GeV and 1.3\,TeV \citep{HESS:2008:PG1553VLT}. The MAGIC
2005-2006 spectra are consistent with this having a power-law photon
index of $\Gamma_{VHE}$=4.21\,$\pm$\,0.25 between $\sim$90 and
500\,GeV \citep{MAGIC:2007:PG1553Detection}. No evidence for
variability of the spectral index, to within the measurement
uncertainties, is seen in the VHE measurements. The H.E.S.S. and MAGIC
integral flux levels are consistent from April to August 2005 and from
April to July 2006 with a mean flux of $I_{VHE}$ (E$>200$\,GeV) =
2.0\,$\pm$\,0.8 $\times$ 10$^{-11}$\,cm$^{-2}$\,s$^{-1}$. Although the
spectrum remained unchanged, the flux detected by MAGIC between
January and April 2006 was lower than the preceding and proceeding
measurements at $I_{VHE}$ ($E\,>200$\,GeV) = 0.6\,$\pm$\,0.2 $\times$
10$^{-11}$\,cm$^{-2}$\,s$^{-1}$. When the systematic uncertainties on
the fluxes are taken into account, this flux is marginally
inconsistent with the VHE fluxes detected during the H.E.S.S. and
other MAGIC observations suggesting that the PG\,1553+113 flux varied
by up to a factor of three on monthly timescales in 2006. PG\,1553+113
has the steepest spectrum of all of the sources detected in the VHE
regime, which makes it a promising target for the \FermiLAT because
extrapolating down to the \Fermi energy range would make this an
extremely bright source unless a dramatic spectral break occurs at
energies below $\sim$100\,GeV. EGRET's non-detection could be
interpreted in this context or as the result of PG\,1553+113 being in
a lower emission state at that time than during the VHE
observations. PG\,1553+113 is in the \FermiLAT bright AGN source list
(LBAS; \citealp{Fermi:2009:LBAS}), the high-confidence AGN
associations from the first three months of \Fermi data. Listed as
0FGL\,J1555.8+1110 with a flux of $I_{LBAS}$
($E$\,$>$\,100\,MeV)\,=\,8.0\,$\pm$\,1.0 $\times$
10$^{-8}$\,cm$^{-2}$\,s$^{-1}$ and a photon index of
$\Gamma_{LBAS}$\,=\,1.70\,$\pm$\,0.06, it has one of the hardest
spectra of the 106 AGN that comprise this list. Indeed, if only the
AGN with a significant detection across the entire \Fermi bandpass are
considered, its spectrum is the hardest of those in the LBAS. This
combination of a very soft VHE spectrum and a very hard HE spectrum
means that PG\,1553+113 has a significant spectral break in the
gamma-ray regime.

Despite significant efforts, the redshift of PG\,1553+113 remains
unknown. Its measurement is of great interest both for a better
understanding of its SED, in particular since it is an extreme BL Lac
and therefore has a very hard synchrotron emission spectrum, and also
for studying EBL effects; some redshift estimates make it the most
distant VHE source detected to date at $z<$0.78
\citep{Sbarufatti:2005:Redshifts}, which, if shown to be the case,
could imply an EBL density close to the minimum allowed by galaxy
counts. The results of redshift studies undertaken to date are
summarized in Section~\ref{SEC:Z} and the archival gamma-ray data
together with the \Fermi data from PG\,1553+113 are used to place
constraints on its redshift.

\section{\Fermi Observations of PG\,1553+113}
\label{SEC:FERMI}

The characteristics and performance of the LAT Instrument on the
{{Fermi}} Gamma-ray Space Telescope are described in detail by
\citet{Atwood:2009:LAT}. Presented here is the analysis of the
\FermiLAT data from a region of 10$^{\circ}$ radius centred on
PG\,1553+113 from 4 August 2008 to 22 February 2009 (MJD
54682.7-54884.2)\footnote{Mission Elapsed Time 239,557,414 to
  256,966,310 seconds.}. These data were analyzed using the standard
\Fermi analysis software\footnote{\texttt{ScienceTools v9r10} with
  instrument response function (IRF) \texttt{P6\_V3\_DIFFUSE}
  \citep{Atwood:2009:LAT}.}. Events with zenith angles $<
105^\circ$ were selected from the so-called ``diffuse class", those
events having the highest probability of being a photon. Only events
with energies greater than 200\,MeV were used in this analysis. The
extragalactic diffuse gamma-ray emission together with the residual
instrumental background was modeled as a simple power-law while the
galactic diffuse was modeled with {\it{GALPROP}}\footnote{The
  background file {\texttt{mapcube\_54\_59Xvarh7S.fits}} was used.}
(\citealp{Strong:2004:GALPROP-letter},
\citealp{Strong:2004:GALPROP-long}).

There are two other \Fermi sources within 10$^{\circ}$ of
PG\,1553+113. These sources, lying at angular separations of
1.8$^{\circ}$ and 5.5$^{\circ}$ from PG\,1553+113 were modeled in our
analysis so that they could be subtracted out along with the
backgrounds described above. The nearest source to PG\,1553+113,
0FGL\,J1553.4+1255, is located at R.A. of $\alpha_{J2000}$ =
15h53m28.2s and declination of $\delta_{J2000}$ = +12d55m20.3s and is
thus spatially coincident with the quasar QSO\,B1551+1305
($z$=1.29). It is detected up to approximately 10\,GeV with an
integral flux ($E>$200\,MeV) of 5.67\,$\pm$\,0.38 $\times$
10$^{-8}$\,cm$^{-2}$\,s$^{-1}$ and photon index of
2.26\,$\pm$\,0.05. There is evidence for variability in its photon
index and flux but neither are correlated with the signal detected
from PG\,1553+113. The LAT has a point spread function (PSF),
$\theta_{68}$, that decreases with increasing energy\footnote{The
  accurate parameterisation of the LAT PSF to be used for science
  analysis is described by the instrument IRF. The following
  simplified, acceptance-averaged approximation for the 68\%
  containment angle might be useful as an illustration of the PSF
  energy dependence:
  ${\theta}_{68}\,{\simeq}\,\sqrt{(0.8^{\circ}\,\times\,E_{\mathrm{GeV}}^{-0.8})^2+(0.07^{\circ})^2}$.}.
For energies above 360\,MeV, the PSF of the LAT is smaller than
1.8$^{\circ}$, the angular separation between PG\,1553+113 and
0FGL\,J1553.4+1255. Therefore, for almost all of the energy range over
which the PG\,1553+113 data are analyzed here
($E$\,=\,200\,MeV\,-\,400\,GeV), the PSF of the LAT is such that the
PG\,1553+113 data are not significantly contaminated by the signal
from 0FGL\,J1553.4+1255. To ensure that this was the case, we
performed an analysis of the PG\,1553+113 data with energy
$E>$\,400\,MeV. The results obtained from this analysis are entirely
consistent with those obtained when the full energy range
($E$\,$>$\,200\,MeV) is considered.

The other source within 10$^\circ$ of PG\,1553+113 is not in the
LBAS. It is located at a R.A. of $\alpha_{J2000}$ = 16h07m40.5s and
declination of $\delta_{J2000}$ = +15d50m36.2s and was detected up to
approximately 10\,GeV with an integral flux ($E>$200\,MeV) of
2.06\,$\pm$\,0.28 $\times$ 10$^{-8}$\,cm$^{-2}$\,s$^{-1}$ and photon
index of 2.31\,$\pm$\,0.11. Its angular separation of 5.5$^\circ$ from
PG\,1553+113 is well in excess of the width of the \Fermi PSF over the
full energy range.

In the \Fermi data analysed here, PG\,1553+113 was detected with a
statistical significance of approximately 49$\sigma$ (\texttt{gtlike}
calculated a test significance\footnote{The test significance is
  defined as -2$\ln$($\Lambda$), where $\Lambda$ is the likelihood
  ratio for the null hypothesis and the assumed source model; see
  \citet{Fermi:2009:BrightSourceList} for a full description.}  of
2407) and an integral flux $I$\,($E$\,$>$\,200\,MeV) of
5.00\,$\pm$\,0.31 $\times$ 10$^{-8}$\,cm$^{-2}$\,s$^{-1}$. The most
energetic photon in the \Fermi data is at 157\,GeV (at an angular
separation of 0.04$^\circ$ from the source location, i.e., well within
$\theta_{68}$ for that energy). The \Fermi data are well described by
a power-law such that the differential photon flux, $F(E)$, is given
by:

\begin{equation}
\label{EQ:dNdE}
  F(E)\,=\,\frac{dN}{dE}\,=\,F_0\,\left(\frac{E}{E_0}\right)^{-\Gamma}
\end{equation}

\noindent where $F_0$ is the differential flux at energy, $E_0$ and
$\Gamma$ is the photon index. For each energy, $E$, the uncertainty
contours, called the butterfly, are defined such that the differential
flux satisfies:

\begin{equation}
\label{EQ:dFF}
\frac{{\Delta}F^2}{F^2}\,=\,\left(\frac{{\Delta}F_0}{F_0}\right)^2\,+\,\textrm{ln}^2\left(\frac{E}{E_0}\right)\Delta\Gamma^2\,-\,\frac{2}{F_0}\,\textrm{cov}(F_0,\Gamma)\,\textrm{ln}\left(\frac{E}{E_0}\right)
\end{equation}

\noindent where cov($F_0,\Gamma$) is the covariance term, returned by
the \texttt{MINUIT} minimization and error analysis function called by
the \Fermi likelihood analysis tool, \texttt{gtlike}, and ${\Delta}F$,
${\Delta}F_0$ and ${\Delta}\Gamma$ are the statistical uncertainties
on the $F$, $F_0$ and $\Gamma$, respectively, at energy, $E$.
Equation~\ref{EQ:dFF} reaches a minimum at the decorrelation energy,
$E_{dec}$, where:

\begin{equation}
\label{EQ:Ed}
E_{dec}\,=\,E_0\,\textrm{exp}\left(\frac{\textrm{cov}(F_0,\,\Gamma)}{F_0\Delta\Gamma^2}\right),\,F_{dec}\,\equiv\,F(E_{dec})
\end{equation}

\begin{figure}
\resizebox*{0.9\textwidth}{!}{\includegraphics[draft=false]{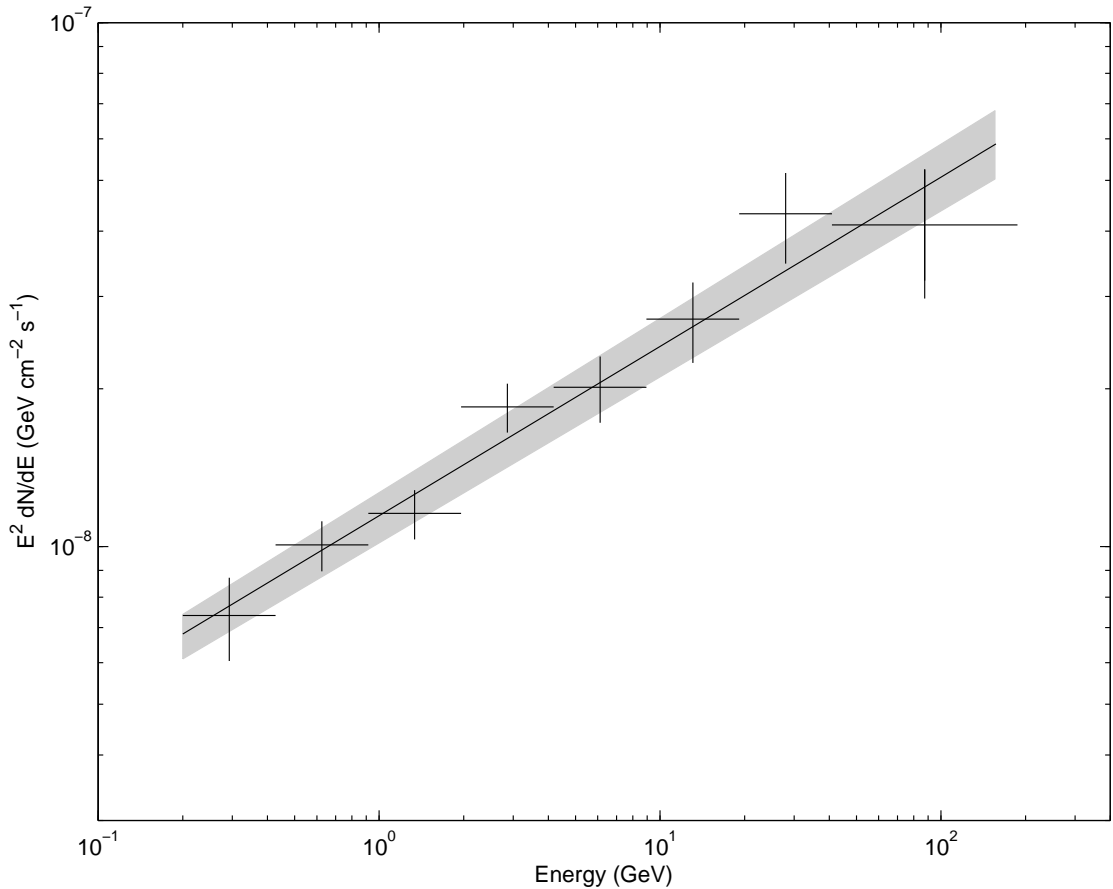}}\\%
\caption{\label{FIG:FermiSpectrum}The differential spectrum from
  PG\,1553+113 as measured by \Fermi. The solid line shows the fit of
  a power-law to the overall spectrum derived from all of the data
  with energy $E>$ 200\,MeV. The data-points (crosses) indicate the
  flux measured in each of the eight energy bins indicated by the
  extent of their horizontal lines, when the data in these energy
  ranges were analyzed with the photon index fixed at the value
  derived from the entire dataset. The grey shaded area shows the
  extent of the \Fermi 68\% confidence band.}
\end{figure}

\noindent For the PG\,1553+113 data analyzed here, we find a
differential flux of $F_{dec}$\,=\,2.60\,$\pm$\,0.18 $\times$
$10^{-9}$ cm$^{-2}$ s$^{-1}$ GeV$^{-1}$ with $E_{dec}$\,=\,2.4\,GeV
and a photon index of $\Gamma$\,=\,1.68\,$\pm$\,0.03. The differential
energy spectrum with the butterfly is shown in
Figure~\ref{FIG:FermiSpectrum}. Also plotted are the fluxes calculated
when the data were analyzed in eight independent energy bins fixing
the photon index in each bin to that derived from the entire dataset,
since the preferred fit to this was a power-law. These fluxes are
listed in Table~\ref{TAB:FermiSpectrum}. To estimate the systematic
uncertainties on the PG\,1553+113 flux and photon index, the
PG\,1553+113 data were re-analyzed using two new sets of instrument
response functions (IRFs; \citealp{Atwood:2009:LAT}) that were created
by propagating both extremes of the uncertainties on IRF
\texttt{P6\_V3\_DIFFUSE}. From this analysis, we estimate the
systematic uncertainties to be at the level of 2\% on the photon index
($\Gamma$\,=\,1.68\,$\pm$\,0.03$_{stat}$\,$^{+0.02}_{-0.04}$$_{syst}$)
and 3\% on the differential flux
($F_{dec}$\,=\,2.60\,$\pm$\,0.18$_{stat}$\,$^{+0.09}_{-0.08}$$_{syst}$
$\times$ $10^{-9}$ cm$^{-2}$ s$^{-1}$ GeV$^{-1}$).

\begin{figure}
\resizebox*{0.9\textwidth}{!}{\includegraphics[draft=false]{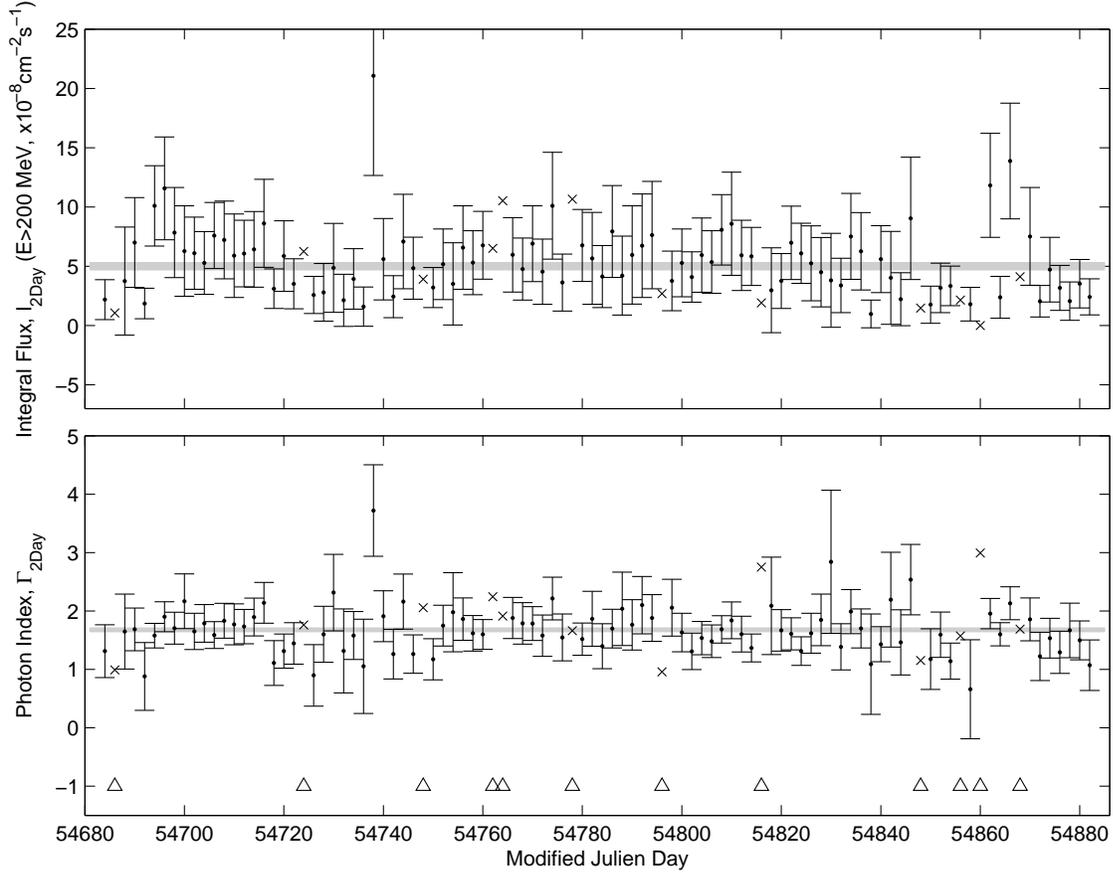}}\\%
\caption{\label{FIG:FermiLightcurve}The lightcurve for PG\,1553+113
  measured by \Fermi between MJD 54683 and 54883 (2008-08-05 -
  2009-02-21). The data are binned in 2-day bins. The top panel shows
  the integral flux, $I_{2Day}$ ($E\,>$\,200\,MeV), for each 2-day
  bin. The bottom panel shows the power-law photon index,
  $\Gamma_{2Day}$, for each 2-day bin. The integral flux and photon
  index, with their uncertainties, calculated by \texttt{gtlike} for
  the entire dataset are shown by the shaded horizontal bands. For
  twelve of the timebins, indicated by the triangles in the lower
  panel, the analysis failed to converge with reasonable error bars
  and therefore, these data were excluded from the variability
  analysis and the points are shown as ``$\times$'' symbols.}
\end{figure}

The PG\,1553+113 200-day lightcurves for the integral flux,
$I_{2Day}$\,($E>$200\,MeV), and the photon index, $\Gamma_{2Day}$,
with 2-day binning are shown in Figure~\ref{FIG:FermiLightcurve}. This
timescale was chosen so that a detection with a statistical
significance of at least 3$\sigma$ was obtained in most bins. These
are the longest continuous lightcurves ever derived for this
source. The $\chi^2$ probability of the data being fit by a constant
are 0.54 and 0.99, respectively, for the integral fluxes and the
photon indices. The normalised excess variance
\citep{Vaughan:2003:ExcVar} is ${\sigma}^2_{NXS}$ = $-$2.7 $\times$
10$^{-2}$ for the flux lightcurve and ${\sigma}^2_{NXS}$ = $-$1.6
$\times$ 10$^{-2}$ for the photon index lightcurve\footnote{It is
  possible to arrive at a negative value for ${\sigma}^2_{NXS}$ when
  there is low intrinsic variance and/or the measurement uncertainties
  are overestimated.}. Both of these calculations suggest that, to
within measurement uncertainties, the flux and photon index from
PG\,1553+113 were constant during these observations. Assuming this to
be the case, an upper limit on the intrinsic variability that could be
present in the flux lightcurve was derived as follows. An ensemble of
lightcurves, each with a different level of intrinsic Gaussian noise
added, was simulated. For each level of intrinsic variability, $v_i$,
1000 lightcurves, each the same length, $L$, as the measured
lightcurve were generated. Each flux point in these lightcurves,
($I_{sim}$)$_{ij}\,\pm\,$(${\Delta}I_{2Day}$)$_j$ ($1 \leq j \leq L$),
was drawn from a Gaussian distribution centred on the mean of the
measured lightcurve with width of
$\sqrt{\left({\Delta}I_{2Day}^2\right)_j+v_i^2}$, where
(${\Delta}I_{2Day}$)$_j$ is the measurement uncertainty for the $j$th
measured flux.  For each of these simulated lightcurves, the excess
variance was calculated, providing a distribution of 1000 simulated
excess variance measurements for each level of added intrinsic
variability. For each of these distributions, the value above which
95\% of the excess variances lay was found. With these data, upper
limits were calculated by constructing a Neyman confidence belt. It
was found that, at the 95\% confidence level, the upper limit on the
normalised intrinsic variance from the PG\,1553+113 flux is $v<$ 7.8
$\times$ 10$^{-2}$. The same procedure was applied to the photon index
lightcurve and an upper limit of $v<$ 1.4 $\times$ 10$^{-2}$ at the
95\% confidence level was derived for the normalised intrinsic
variance on the PG\,1553+113 photon index.

\section{Using the Gamma-ray Data to Constrain the Redshift of PG\,1553+113}
\label{SEC:Z}

For many years PG\,1553+113 was thought to lie at a redshift of
$z$=0.36 \citep{Miller:1983:PG1553z}. This redshift estimate was based
on a spurious emission line in its spectrum, misidentified as
Lyman-$\alpha$ \citep{Falomo:1990:PG1553}. Subsequent observations
have failed to reveal any spectral features
(\citealp{Falomo:1990:PG1553}; \citealp{Falomo:1994:SpecBlazars};
\citealp{Carangelo:2003:SpecBLLacs}). Since its detection in the VHE
band, a number of dedicated observing campaigns and reanalyses of
Hubble Space Telescope (HST) images have been undertaken to determine
its redshift but, despite these efforts, the redshift of PG\,1553+113
remains unknown, (\citealp{Sbarufatti:2005:Redshifts};
\citealp{Sbarufatti:2006:ESO-12Redshifts};
\citealp{Treves:2007:PG1553z};
\citealp{HESS:2008:PG1553VLT}). Constraints, some of them
contradictory, have been placed on its redshift using a variety of
techniques. These fall into two categories --- spectral and imaging
observations in optical to ultraviolet wavebands and VHE gamma-ray
spectral observations.

The host galaxy of PG\,1553+113 was neither resolved by
\citet{Hutchings:1992:OpticalQSOs} using the High-Resolution camera on
the Canada France Hawaii Telescope nor by
\citet{Scarpa:2000:HSTBLLacs} who observed it as part of a survey of
110 BL Lacs using HST. The host galaxies of 69 out of these were
resolved by HST, including almost all objects with $z<0.5$.
\citet{Sbarufatti:2005:Redshifts} analyzed these HST data and showed
that the host galaxy luminosity is encompassed in a relatively narrow
range for the BL Lacs resolved with HST, thus concluding that BL Lac
host galaxies typically have absolute magnitude of
$M_R\,=\,-22.8$. They used this assumption to derive lower limits on
the redshifts for those objects lying at unknown distances for which
no host galaxy could be resolved, thus arriving at a lower limit of
$z>0.78$ for PG\,1553+113, the brightest such object in the
survey. Using a similar strategy, \citet{Carangelo:2003:SpecBLLacs}
set a lower limit of $z>0.3$ on the redshift of PG\,1553+113 using
preliminary results from observations with the ESO $3.6$\,m
telescope. \citet{Treves:2007:PG1553z} reanalyzed the HST image of
PG\,1553+113 and, again making the assumption that its host galaxy is
typical of BL Lacs, placed a lower limit of $z\geq0.25$ on the
redshift. In comparing their results with those from VHE observations,
they concluded that its redshift is in the range
$z=0.3-0.4$. \citet{Sbarufatti:2006:ESO-12Redshifts} used the ESO VLT
to measure the optical spectra of a number of BL Lacs including
PG\,1553+113. Combining the fact that no spectral features were
resolved with their knowledge of the sensitivity of the VLT to such
features, they were able to derive a lower limit of $z>0.09$ on the
redshift PG\,1553+113, under the assumption that its host galaxy is
typical.

All of the results above, however, rely on the result of
\citet{Sbarufatti:2005:Redshifts}, which states that BL Lac host
galaxies have a very small dispersion in their absolute magnitudes.
The results of a study by \citet{ODowd:2005:Evolution} casts doubt on
the validity of this assumption for objects with $z>0.5$ because they
find evidence for strong evolution in the host galaxies of BL Lacs in
the redshift range of $z=0.5-2.5$.

Gamma-ray data can be used indirectly to estimate redshifts provided
that a few necessary assumptions are valid. The VHE spectra of
extragalactic sources whose redshifts are known can be used to probe
the density of the EBL because the high-energy gamma rays produce
e$^+$e$^-$ pairs with the EBL photons thus introducing a
redshift-dependent absorption feature on the spectra observed in the
VHE regime (\citealp{Costamante:2004:4BLLacsEBL};
\citealp{Dwek:2005:EBL}; \citealp{Stecker:2007:EBL};
\citealp{Krennrich:2008:EBL}). Conversely, in cases such as that of
PG\,1553+113, a firmly established VHE source
(\citealp{HESS:2006:PG1553Detection};
\citealp{MAGIC:2007:PG1553Detection}; \citealp{HESS:2008:PG1553VLT};
\citealp{MAGIC:2009:PG1553MWL}) lying at an unknown distance, the
measured spectrum can be used in combination with assumptions about
the density of the EBL and the intrinsic VHE spectrum to put limits on
its redshift. By assuming a minimal level of EBL
\citep{Primack:2001:EBL} and that the intrinsic spectrum of
PG\,1553+113 is unlikely to have a photon index harder than
${\Gamma}_{int}$ = 1.5, the limit from shock acceleration models
\citep{HESS:2006:EBL-nature}, \citet{HESS:2006:PG1553Detection}
derived an upper limit of $z<0.74$ on its redshift. We note that none
of the \Fermi LBAS sources that have significant detections across the
entire \Fermi bandpass have photon indices harder than 1.5, providing
additional support for the validity of this assumption. When these
data were reanalysed and combined with subsequent H.E.S.S. data, a
refined upper limit of $z<0.69$ was calculated
\citep{HESS:2008:PG1553VLT}. A similar procedure was adopted by
\citet{MAGIC:2007:PG1553Detection} resulting in an upper limit of
$z<0.78$. \citet{Mazin:2007:PG1553} combined all of the existing VHE
gamma-ray data from PG\,1553+113 and, by assuming that a break in the
intrinsic spectrum should be visible if the source lies above
$z=0.42$, as well as a limit on the hardness of the intrinsic spectrum
${\Gamma}_{int}$ = 1.5, placed an upper limit of $z<0.42$ on the
redshift of PG\,1553+113.

\begin{figure}
\resizebox*{0.9\textwidth}{!}{\includegraphics[draft=false]{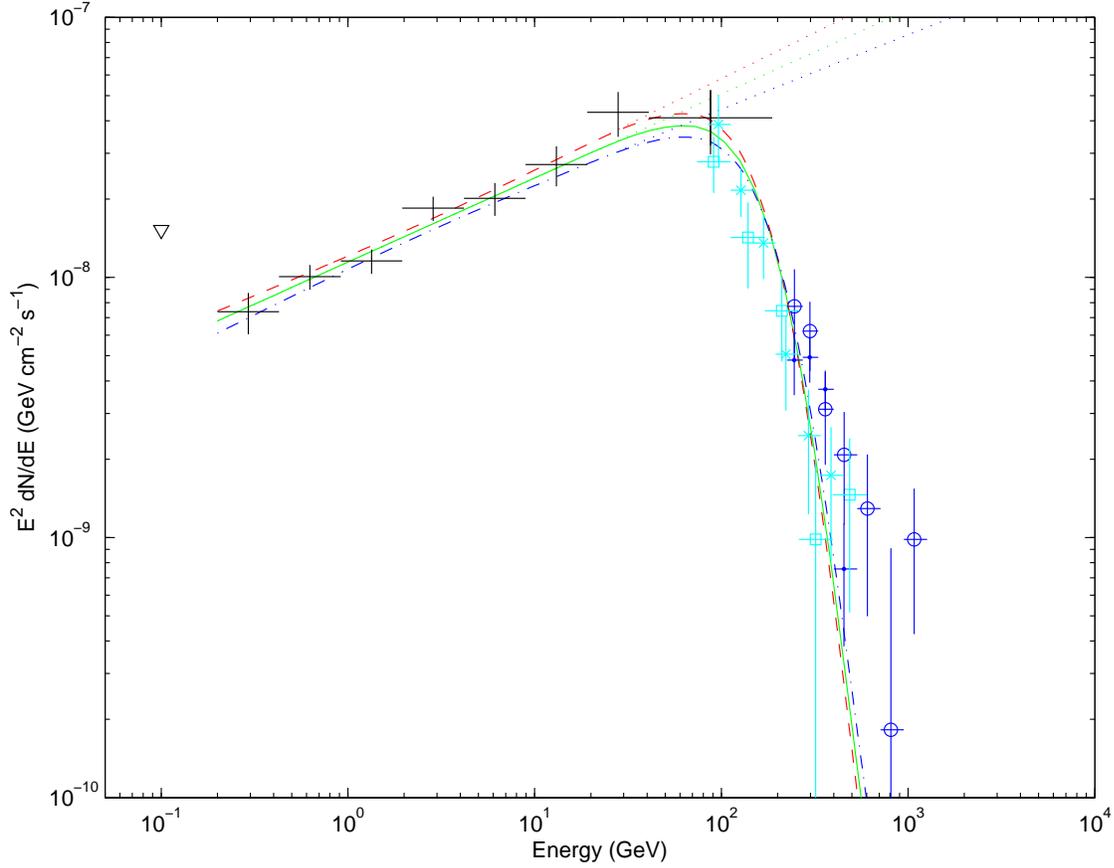}}\\%
\caption{\label{FIG:SED-EBL}The spectral energy distribution for the
  gamma-ray data. The individual datasets are described in the
  text. The \Fermi datapoints are shown as black crosses. The EGRET
  upper limit for emission above 100\,MeV is shown as a black
  triangle. The H.E.S.S. data combined from May \& August 2005 are
  shown as blue dots and those combined from April \& July 2006 as
  blue open circles. The MAGIC data combined from April \& May 2005
  and from January to April 2006 are shown as cyan x's while those
  from July 2006 are shown as cyan open squares. The green solid line
  shows the power-law fit to the \Fermi data, extended to higher
  energies with the level of EBL that best fitted the VHE data, which
  corresponds to the EBL integrated to a redshift of $z=0.75$. The
  green dotted line shows the extension of the \Fermi power-law to
  higher energies without absorption. The upper and lower 68\%
  uncertainty-contours for the \Fermi data are shown as red dashed and
  blue dash-dotted lines, respectively. Each of them were also
  extended with the level of EBL that best fitted the VHE data, which
  corresponded to a redshift of $z=0.79$ for the upper contour and to
  a redshift of $z=0.70$ for the lower contour. Their unabsorbed
  extensions to higher energies are shown as red and blue dotted
  lines.}
\end{figure}

The gamma-ray SED measured by \Fermic, H.E.S.S. and MAGIC are plotted
together in Figure~\ref{FIG:SED-EBL}. The VHE data are
non-simultaneous with those from \Fermi and comprise two
H.E.S.S. datasets and two MAGIC datasets
(\citealp{HESS:2006:PG1553Detection}; \citealp{HESS:2008:PG1553VLT};
\citealp{MAGIC:2007:PG1553Detection};
\citealp{MAGIC:2009:PG1553MWL}). The VHE spectra recorded by
H.E.S.S. and by MAGIC were consistent during all
observations. Although one of the MAGIC datasets showed evidence for a
change in the mean flux level from PG\,1553+113 on month timescales,
no evidence for strong variability or day-scale (or shorter) VHE flux
variability is seen, within the measurement uncertainties, in any of
the datasets. As described in Section~\ref{SEC:FERMI} no flux or
spectral variability was detected in the \Fermi dataset. We therefore
make the assumption here that the source was in a non-flaring state
during all of the gamma-ray observations. Furthermore, since the
highest energy \Fermi datapoints, which overlap with the energy range
covered by MAGIC, show a flux level consistent with that measured by
MAGIC, we assume that PG\,1553+113 was in a similar flux and spectral
state at all epochs plotted in Figure~\ref{FIG:SED-EBL}. We note that
in both the optical and X-ray regimes, the spectral properties of
PG\,1553+113 were not seen to change significantly even when its flux
level changed.

The spectrum measured by \Fermi is best-fit by a simple power law. We
make the assumption that any departures from this power-law spectrum
up to 1\,TeV are dominated by absorption of gamma rays by the EBL and
use the parameterisation of \citet{Franceschini:2008:EBL}, which
includes evolutionary effects, to find the level of EBL, and therefore
the redshift, which best fits the measured data. By absorbing the
extrapolated \Fermi spectrum with EBL corresponding to redshifts from
$z=0.01-3.00$, in steps of $z=0.01$, we find that a redshift of
$z=0.75$ gives the best ${\chi}^2$ fit to the measured VHE data for
the EBL model of \citet{Franceschini:2008:EBL}. When the \Fermi 68\%
uncertainty-contours, derived from Equation~\ref{EQ:Ed}, are subjected
to the same fitting procedure as the power-law spectrum we obtain the
statistical error on the redshift determination for the
\citet{Franceschini:2008:EBL} EBL giving us a range from
$z=0.70-0.79$, as illustrated in Figure~\ref{FIG:SED-EBL}. Since the
model of \citet{Franceschini:2008:EBL} provides the lowest level for
the EBL over the range of interest here, the redshift derived from it
should be considered an upper limit. Due to the fact that we performed
this calculation for just one parameterization of the EBL we do not
estimate the systematic uncertainty associated with absorbing the
\Fermi spectrum with different EBL models and we discuss this further
in Section~\ref{SEC:CONCLUSION}. We note that the systematic
uncertainty on the energy scale, which is on the order of $10-15$\%
for the VHE gamma-ray telescopes (\citealp{HESS:2006:Crab};
\citealp{MAGIC:2008:Crab}), was not taken into account in these
calculations.

\begin{deluxetable}{lccccc}
\tablewidth{0pt} \tablecaption{\label{TAB:X-RAY}The flux and spectral
  parameters from X-ray observations (2-10\,keV, unless otherwise
  noted) of PG\,1553+113. The spectral parameters for a log-parabola
  fit are shown with the differential X-ray Flux, $F_{X}$(E) $\propto$
  E$^{-a -b(\log(E))}$ (cm$^{-2}$\,s$^{-1}$\,keV$^{-1}$). In
  cases where no value is given for $b$, the best-fit power-law
  spectrum is shown with $F_{X}$($E$) $\propto$ $E^{-a}$
  (cm$^{-2}$\,s$^{-1}$\,keV$^{-1}$). The abbreviated references
  correspond to the following: [Don01] - \citet{Donato:2001:HardXray};
  [Don05] - \citet{Donato:2005:BeppoSAXBlazarCat}; [Per05] -
  \citet{Perlman:2005:XraySpectra}; [Ost06] -
  \citet{Osterman:2006:PG1553mwl}; [Tra07] -
  \citet{Tramacere:2007:SwiftTeV}; [Rei08] -
  \citet{Reimer:2008:PG1553Suzaku}; [Mas08] -
  \citet{Massaro:2008:Xray11years}.}

\tablehead{           &                                          & \colhead{Flux (2-10\,keV)}                         & \multicolumn{3}{c}{Spectral}       \\
                      & \colhead{Observation}                    & \colhead{($\times$ 10$^{-11}$\,erg\,cm$^{-2}$\,s$^{-1}$)} & \multicolumn{3}{c}{Parameters}       \\
\colhead{Observatory} & \colhead{Date(s)}                        & \colhead{[Ref.]}                                   & \colhead{$a$}&\colhead{$b$}&\colhead{[Ref.]}}
\startdata

\ASCA                 & 1995-08-16                               & 2.9 [Don01]          & 2.47 & ---    & [Don01] \\ 
\BeppoSAX             & 1998-02-05                               & 1.4 [Don05]          & 2.17 & 0.63   & [Mas08] \\ %
\XMMNewton            & 2001-09-06                               & 3.5 [Per05]          & 2.09 & ---    & [Mas08] \\ %
\RXTE                 & 2003-04-22 - 2003-05-12\tablenotemark{a} & 0.7 [Ost06]          & 2.37 & ---    & [Ost06] \\
                      & 2003-04-22 - 2003-05-28\tablenotemark{b} & 0.5 [Ost06]          & 2.60 & ---    & [Ost06] \\ 
                      & 2003-04-26\tablenotemark{c}              & 0.3 [Ost06]          & 3.19 & ---    & [Ost06] \\ 
                      & 2003-05-11\tablenotemark{d}              & 1.2 [Ost06]          & 2.26 & ---    & [Ost06] \\ 
\SwiftXRT             & 2005-04-20                               & 2.1 [Tra07]          & 2.21 & 0.36   & [Mas08] \\ %
                      & 2005-10-06                               & 6.9 [Tra07]          & 2.14 & 0.24   & [Mas08] \\ 
                      & 2005-10-08                               & 6.7 [Tra07]          & 2.11 & 0.23   & [Mas08] \\ 
\Suzaku               & 2006-07-24 - 2006-07-25                  & 3.5 [Rei08]          & 2.19 & 0.26\tablenotemark{e} & [Rei08] \\ 
\enddata

\tablenotetext{a}{The mean flux and spectral index during the whole campaign.}
\tablenotetext{b}{The mean flux and spectral index prior to the flare.}
\tablenotetext{c}{The minimum flux recorded during the campaign and the spectral index at that time.}
\tablenotetext{d}{The maximum flux recorded during the campaign and the spectral index at that time.}
\tablenotetext{e}{The \Suzaku spectrum was measured between 0.3-30\,keV.}

\end{deluxetable}

\begin{deluxetable}{cc}
\tablewidth{0pt} \tablecaption{\label{TAB:FermiSpectrum}The
  differential flux measured by the \Fermi LAT in each energy
  bin.}

\tablehead{\colhead{Energy Range} & \colhead{Flux}                            \\
\colhead{(GeV)}                   & \colhead{(cm$^{-2}$\,s$^{-1}$\,GeV$^{-1}$)}}
\startdata

0.20 -   0.43                     & 8.63\,$\pm$\,1.56 $\times$ 10$^{-8}$      \\
0.43 -   0.92                     & 2.57\,$\pm$\,0.28 $\times$ 10$^{-8}$      \\
0.92 -   1.96                     & 6.46\,$\pm$\,0.70 $\times$ 10$^{-9}$      \\
1.96 -   4.18                     & 2.26\,$\pm$\,0.24 $\times$ 10$^{-9}$      \\
4.18 -   8.94                     & 5.38\,$\pm$\,0.78 $\times$ 10$^{-10}$     \\
8.94 -   19.13                    & 1.59\,$\pm$\,0.28 $\times$ 10$^{-10}$     \\
19.13 -  40.90                    & 5.52\,$\pm$\,1.08 $\times$ 10$^{-11}$     \\
40.90 -  187.05                   & 5.37\,$^{+1.49}_{-1.16}$\tablenotemark{\dagger} $\times$ 10$^{-12}$     \\

\enddata

\tablenotetext{\dagger}{The uncertainty in the highest energy bin is found to be asymmetric.}

\end{deluxetable}

\begin{deluxetable}{lccccc}
\tablewidth{0pt} \tablecaption{\label{TAB:SED-PARAMS}The parameters of
  the electron distribution in the SSC model, as described in the
  text, used to fit the PG\,1553+113 data. The dates of the
  observations are listed at the top of each column. The colour of the
  line used to show each of the models in Figure~\ref{FIG:SSC} is
  listed in parenthesis beneath each column's header. For each
  dataset, the minimal Lorentz factor is ${\gamma}_{min}\,=\,1.00$ and
  we find an equipartition factor of 0.2 (ratio of the energy density
  in the magnetic field to that in the particles of the jet).}

\tablehead{\colhead{Model}         & \colhead{\RXTE}       & \colhead{\SwiftXRT}        & \colhead{\Suzaku}       & \colhead{\SwiftXRT}        & \colhead{No X-rays}        \\
\colhead{Parameter}                & \colhead{(Blue)}      & \colhead{(Yellow)}         & \colhead{(Green)}       & \colhead{(Magenta)}        & \colhead{(Black)}        }
\startdata

                                   & Apr/May 2003          & Oct 2005                   & Jul 2006                & Mar 2009                   & --- \\
\\
$p_{1}$                            & 1.70                  & 1.70                       & 1.70                    & 1.70                       & 1.70                       \\
$p_{2}$                            & 3.00                  & 3.00                       & 2.70                    & 3.00                       & 3.00                       \\
$p_{3}$                            & 4.10                  & 4.10                       & 3.90                    & 4.10                       & ---                        \\
${\gamma}_{max}$ $\times$ 10$^{6}$ & 3.16                  & 3.16                       & 3.16                    & 3.16                       & 0.20                       \\
${\gamma}_1$ $\times$ 10$^{4}$     & 6.59                  & 6.59                       & 6.59                    & 5.07                       & 5.71                       \\
${\gamma}_2$ $\times$ 10$^{4}$     & 7.61                  & 22.8                       & 7.61                    & 6.59                       & ---                        \\
$D_{tot}$ $\times$ 10$^{54}$       & 4.00                  & 3.70                       & 3.93                    & 4.28                       & 4.32                       \\

\enddata
\end{deluxetable}

\section{Discussion}
\label{SEC:DISCUSSION}

The \Fermi data presented here allowed us to derive the longest,
continuously sampled lightcurves to date for PG\,1553+113. Its flux in
this energy regime is such that its variability can be probed on
$\sim$day timescales. The combination of the low duty cycle of VHE
instruments and the weak flux from PG\,1553+113 at those energies
means that the timescales accessible with the LAT for this source are
shorter than those accessible at higher energies where longer
integration times than those available on nightly timescales have been
necessary to achieve a significant detection. No evidence for
variability was found in the \Fermi integral flux and photon index
lightcurves. That there is no evidence for variability from
PG\,1553+113 is consistent with observations of other BL Lacs with
\Fermi and the VHE instruments. The recently-released \Fermi LBAS
\citep{Fermi:2009:LBAS} found evidence that BL Lacs are less variable
than the other blazar subclass, the flat spectrum radio quasars
(FSRQs). Using a simple $\chi^2$ test, 70\% of the LBAS FSRQs were
found to be variable compared to 29\% of the LBAS BL Lacs. Also, the
fact that approximately 70\% of the EGRET-detected blazars are not in
the LBAS, with comparable flux thresholds, is a further indication
that high activity in the gamma-ray range is not frequent for a given
source. As the number of BL Lacs detected in the VHE regime increases,
it seems that these objects are not all as variable in the VHE regime
as initial observations of Markarian\,421 and Markarian\,501 suggested
(e.g., \citealp{Buckley:1996;Mrk421Variability},
\citealp{Gaidos:1996:Mrk421Flare}, \citealp{HEGRA:1997:Mrk501},
\citealp{Quinn:1999:Mrk501Var}). Of the 21 BL Lacs now confirmed to be
VHE gamma-ray emitters\footnote{http://tevcat.uchicago.edu; The
  ``Default Catalog'', which contains only those sources that have
  been confirmed as VHE emitters, was used (see TeVCat website for
  more details).}  \citep{Wakely:2008:TeVCat}, 10 (48\%) have, to
date, shown no evidence for strong variability, 4 (19\%) have shown
marginal evidence for variability (i.e., different flux levels on
yearly timescales), and 7 (33\%) show strong evidence for
variability. It should be noted however, that selection effects can
play a role here: the low duty cycles and small fields of views of VHE
instruments limits their ability to perform routine monitoring of a
large sample of sources and, in many cases, BL Lac observations are
triggered by high flux states. \Fermic, on the other hand, with its
large field of view and high duty cycle, can sample the lightcurves of
sources in a more even fashion, in particular over day-scale
timebins. It is, however, less sensitive than VHE instruments to
shorter-term variability of sources with spectral and flux
characteristics typical of BL Lacs detected to date. Threshold
effects, also, can play a role in the detection of variability from a
source. When the falling edge of the source's SED intersects with the
low energy range of the instrument sensitivity (this is the case for
the FSRQs with \Fermi and the BL Lacs with the VHE instruments), a
slight spectral shift from the source can mimic the effects of a large
level of variability.

\subsection{Modeling the Intrinsic Emission of PG\,1553+113}
\label{SEC:MODELING}

\begin{figure}
\resizebox*{0.9\textwidth}{!}{\includegraphics[draft=false]{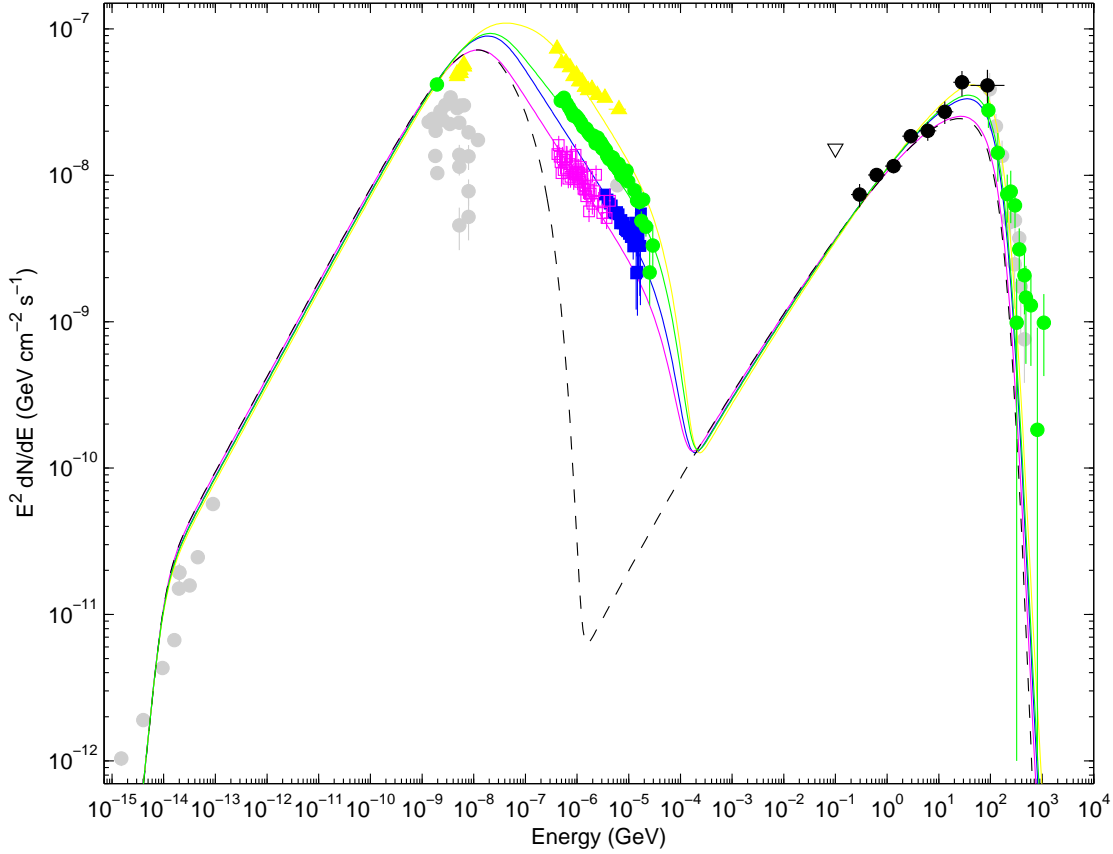}}\\%
\caption{\label{FIG:SSC}The spectral energy distribution for
  PG\,1553+113 fit with an SSC model. The fitting procedure is
  described in the text. The yellow, green, blue and magenta lines
  are, respectively, the SSC model tuned to fit the \Swift \XRT data,
  the \Suzaku data, the \RXTE data and the \Swift \XRT data. The black
  dashed line is the SSC model when the highest energy electrons are
  omitted entirely from the fit. The parameters are described in the
  text and are summarised along with the dates of the observations in
  Table~\ref{TAB:SED-PARAMS}. The X-ray data have been de-absorbed for
  a column density of
  $N_H$\,=\,3.67$\times$10$^{-20}$\,cm$^{-2}$. Apart from 7 hours of
  the H.E.S.S. data (taken in April 2006) the KVA (optical;
  \citealp{Reimer:2008:PG1553Suzaku}), \Suzaku
  \citep{Reimer:2008:PG1553Suzaku}, MAGIC \citep{MAGIC:2009:PG1553MWL}
  and H.E.S.S. \citep{HESS:2008:PG1553VLT} observations were made in
  July 2006 and are thus quasi-simultaneous. They are shown as green
  filled circles. The \SwiftXRT and \UVOT data from October 2005
  \citep{Tramacere:2007:SwiftTeV} are shown as yellow triangles while
  the magenta open squares are \SwiftXRT data (Swift observation ID
  31368001) from March 2009. The \RXTE data from
  \citet{Osterman:2006:PG1553mwl} are shown as blue filled
  squares. The archival data (grey filled circles) come from:
  \citet{Bennett:1986:PG1553SEDned}, \citet{Becker:1991:GHzSources},
  \citet{Gregory:1991:87GB}, \citet{Douglas:1996:TexasRadio},
  \citet{Gorshkov:2003:Radio} (\radioc); \citet{Falomo:1990:PG1553},
  \citet{Urry:2000:PG1553SEDned},
  \citet{Sbarufatti:2006:ESO-12Redshifts}, the Sloan Digital Sky
  Survey \citep{SDSS:2009:SDSS7} via NED,
  \citet{Fox:2006:HighVelClouds}, \citet{Tramacere:2007:SwiftTeV},
  (\opticalUVc); \citet{Donato:2005:BeppoSAXBlazarCat}, (\Xrayc);
  H.E.S.S. \citep{HESS:2008:PG1553VLT} and
  \citet{MAGIC:2007:PG1553Detection}, (\VHEgammarayc).}
\end{figure}

\begin{figure}
\resizebox*{0.9\textwidth}{!}{\includegraphics[draft=false]{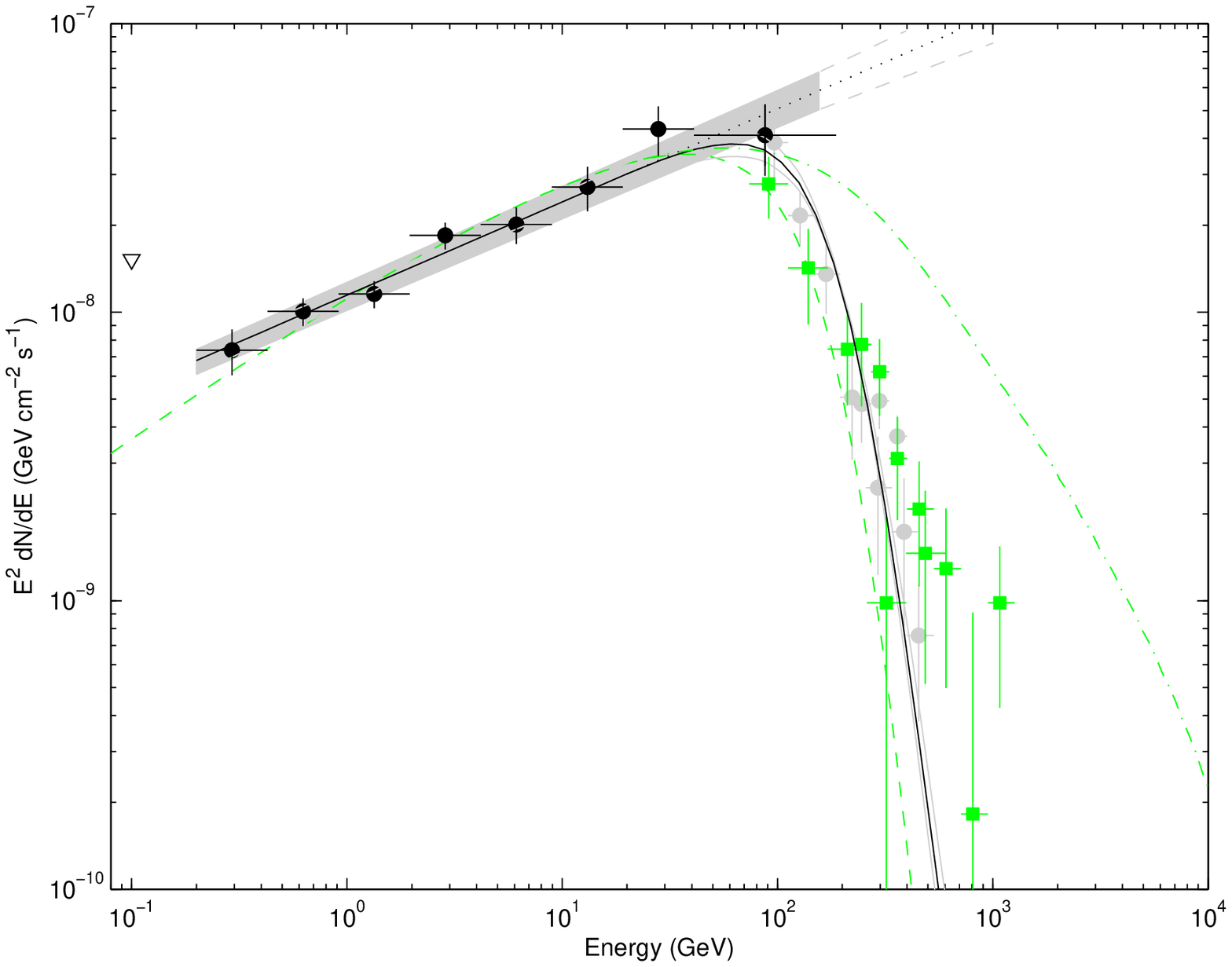}}\\%
\caption{\label{FIG:ZOOM}A zoom on the high-energy portion of the
  spectral energy distribution for PG\,1553+113. The \Fermi datapoints
  are shown by black filled circles; the 2006 H.E.S.S. and MAGIC data
  are shown as green solid squares while the H.E.S.S. and MAGIC data
  from 2005 are shown as grey solid circles. The EGRET upper limit is
  shown as a black triangle. The best SSC model fit for the \Suzaku
  X-ray dataset, with EBL absorption applied, is shown as a green
  dashed line. The green dash-dotted line shows the SSC model before
  absorption for the EBL. The black solid line shows the Fermi
  power-law spectrum absorbed for $z$\,=\,0.75 using the model of
  \cite{Franceschini:2008:EBL} while the dotted black line shows the
  unabsorbed Fermi spectrum. The shaded area shows the \Fermi
  butterfly and the grey dashed lines, its unabsorbed extension to
  higher energies. The grey solid lines show the \Fermi butterfly
  absorbed for the best fit redshift to each edge, as discussed in
  Section~\ref{SEC:Z}.}
\end{figure}

In accordance with the weak observed variability, all of the available
PG\,1553+113 gamma-ray data were combined in turn with each of the
X-ray datasets and, assuming a redshift of $z$=0.75 as found above,
each dataset was modeled with a single-zone synchrotron self-Compton
(SSC) model, that is, a scenario in which one population of electrons
is responsible for the broadband emission, producing synchrotron
radiation in the radio to X-ray regime and upscattering these
synchrotron photons to produce the gamma-ray emission
(\citealp{Band:1985:SSC}; the model employed here is similar to the
one used in \citet{FermiHESS:2009:PKS2155MWL}). Figure~\ref{FIG:SSC}
shows the result of this fitting. The electrons are parameterized as a
three-component power-law, $dn/d\gamma \propto \gamma^{-p_i}$
($i=1-3$), with minimal and maximal Lorentz factors ${\gamma}_{min}$
and ${\gamma}_{max}$, break Lorentz factors of ${\gamma}_{1}$ and
${\gamma}_{2}$ and total electron number of $D_{tot}$. For each of the
X-ray datasets, the SED was modeled by finding the parameters of the
electron distribution that provided a good fit to the shape of the
low-energy component (radio to X-ray), while keeping the same values
for the emitting region radius, $R$, the magnetic field strength, $B$,
and the bulk Doppler factor, $\delta_{bulk}$ throughout. This fitting
procedure resulted in values of $R=1.4\times10^{18}$\,cm,
$B=0.01$\,Gauss and $\delta_{bulk}=32$.  Table~\ref{TAB:SED-PARAMS}
lists the best fit values of the electron parameters for each of the
X-ray datasets. A zoom of the high-energy component of the SED for the
\Fermi data with the dataset that is most simultaneous (that
corresponding to the KVA, \Suzaku and VHE data from July 2006),
without the effects of EBL absorption, is shown in
Figure~\ref{FIG:ZOOM}. It can be seen that, even with the presence of
the intrinsic curvature that is inherent to the SSC model, most of the
curvature in the high-energy portion of the spectrum can be accounted
for by EBL absorption.

We found that this simple, one-zone SSC model provided a reasonably
good fit to the PG\,1553+113 SED. By altering only the distribution of
the electrons that produce the synchrotron emission, a good fit to the
overall SED was found for each of the X-ray flux states and, for all
of these model realisations the VHE component of the SED did not
change significantly: the magnitude of the changes in the SED above
$\sim$200\,GeV are on the order of the VHE statistical measurement
uncertainties. This consistency of spectral shape implies that the
gamma-ray flux could remain consistent with the state seen by
H.E.S.S. and MAGIC during the observations of 2005-2006, even in the
presence of the large changes in the X-ray flux level that have been
detected. Such behaviour was also observed with simultaneous datasets
from PKS\,2155-304 by \citet{FermiHESS:2009:PKS2155MWL}. In a
quiescent state during these observations, no correlation was found
between its X-ray and VHE gamma-ray fluxes. During previous
observations however, when it was in a flaring state in the gamma-ray
regime, there was strong correlation between the X-ray and VHE fluxes
\citep{Costamante:2008:PKS2155}. This behaviour is an indication that
the hard X-ray flux of BL Lacs can change significantly without
resulting in detectable activity in the gamma-ray regime, except for
at the peak of the SED at these energies, i.e., that X-ray variability
can be accompanied by VHE gamma-ray quiescence and a measurable shift
in the spectrum at the peak of the \Fermi energy regime. In such a
scenario, the electrons producing the variable X-ray emission are at
higher energies than those upscattering the bulk of the synchrotron
photons to the VHE gamma-ray regime; the scatterings of the variable
hardest X-rays are suppressed mostly due to Klein-Nishina effects but
also because of the decreasing target photon density at these
energies. This effect is demonstrated for the extreme case in
Figure~\ref{FIG:SSC} where the black dashed curve shows the broadband
spectrum when the high-energy electron component is omitted. It can be
seen that the difference between it and the curves in which the
highest energy electrons are included is still on the order of the
statistical measurement uncertainties in the VHE regime.

In the framework of this model, the X-ray and VHE gamma-ray fluxes
would be correlated during gamma-ray flaring states while during the
more common gamma-ray quiescent states changes in the X-ray flux would
not lead to detectable changes in the gamma-ray flux, except for at
the high-energy peak in the SED. The simple, one-zone SSC model
employed here allows for such a scenario --- accounting for the X-ray
flux variations observed historically while not requiring detectable
changes in the VHE gamma-ray flux state. It provides a good fit to the
VHE gamma-ray data with each of the low-energy datasets, only one of
which, drawn in green in Figure~\ref{FIG:SSC}, is quasi-simultaneous
(the KVA, \Suzaku and all but 7 hours of the VHE data are from July
2006). It remains to be seen whether such behaviour is observed in
general from BL Lacs, an investigation that can be undertaken jointly
by \Fermi and the X-ray and VHE observatories. Correlations between
the arrival of $>$100\,GeV photons in the \Fermi data and increasing
of the hard X-ray flux, in the absence of VHE variability, would be
indicative of the proposed scenario. Such effects would be difficult
to pin down however, due to the low rate of detection of of
$>$100\,GeV photons with \Fermic.

\citet{Vercellone:2004:DutyCycle} studied the duty-cycle of gamma-ray
blazars by comparing the sources detected by EGRET with a sample of
radio-selected candidate gamma-ray blazars. They found that most
blazars have a duty cycle of less than 10\%, meaning that they spend
more than 90\% of their time in a non-flaring gamma-ray state. The
gamma-ray data presented here suggest that PG\,1553+113 was in such a
state in the \Fermi energy range during all of these
observations. Given the increase in sensitivity and duty cycle
afforded us by \Fermic, we should expect to detect many more blazars
in their quiescent states in the $>$200\,MeV energy range.

\section{Conclusion}
\label{SEC:CONCLUSION}

We have combined the \Fermi data from PG\,1553+113 with data from
radio through VHE gamma rays to study its broadband emission. We
demonstrated that a simple, one-zone SSC model provides a reasonably
good fit to the observational results: It accounts for the different
X-ray flux states observed while also allowing the gamma-ray data to
remain approximately constant. More detailed theoretical modeling of
the SED, which naturally might fit the data points better, is beyond
the scope of this paper.

We have used gamma-ray data spanning four orders of magnitude to seek
the EBL column density (as parameterized by
\citet{Franceschini:2008:EBL}) that best fits the measured spectrum of
PG\,1553+113. We find that an EBL integrated to a redshift of
$z=0.75^{+0.04}_{-0.05}$ provides the best fit. Such a high value for
the redshift would make PG\,1553+113 the most distant source to be
detected in the VHE regime, an attribute consistent with it being the
VHE source with the largest spectral break in the gamma-ray regime
($\Gamma_{VHE}$ - $\Gamma_{HE}$ $\sim$ 2.7). Three assumptions were
made in our redshift calculation the first one being that the EBL
behaves in a manner consistent with the model of
\citet{Franceschini:2008:EBL}. Of the many EBL models available in the
literature (see \citet{Finke:2009:EBL} for a recent comparison), that
of \citet{Franceschini:2008:EBL} predicts the lowest level of EBL
across the energy range of interest here; it provides the minimum
level of EBL photons that could exist based on known sources alone and
therefore, in using it, the redshift derived for PG\,1553+113 can be
considered an upper limit. Had one of the other EBL models been used,
the best-fit redshift would have been lower. Indeed, a more in-depth
study of the absorption effects of the EBL on the spectrum of
PG\,1553+133 should consider all of the EBL models available in the
literature. In this way, the systematic effects of using different
realisations of the EBL could be determined; the data points provided
in Table~\ref{TAB:FermiSpectrum} allow for such a study to be
undertaken.

Two additional assumptions were made in our redshift calculation,
firstly, that the emission state of PG\,1553+113 did not change
significantly during the gamma-ray observations, which can be
interpreted as an indication that it was in a low-flux state state at
these energies throughout those observations. Indeed, the quiescent
state is the state in which we are most likely to find a blazar
\citep{Vercellone:2004:DutyCycle}. Secondly, it was assumed that the
power-law spectrum measured by \Fermi does not suffer from significant
intrinsic absorption up to 1\,TeV. Should there be intrinsic
absorption of VHE gamma rays at the source, a lower level of EBL
absorption would be required to best fit the measured VHE spectrum. We
note that, since the \Fermi data show no evidence for spectral
curvature, if it is significant intrinsic absorption that is mostly
responsible for the sharp break in the gamma-ray spectrum, its effects
would have to kick in at energies exactly above those accessible to
\Fermi and, at a significant level, in order to produce the sharp
break observed in the data.

This is the first time that the measurement of the complete HE to VHE
gamma-ray spectrum of a source has been used to constrain its
redshift.  The value derived here is close to the limits derived by
\citet{HESS:2008:PG1553VLT} and \citet{MAGIC:2007:PG1553Detection} and
is higher than that derived by \citet{Mazin:2007:PG1553}. Different
parameterizations of the EBL were used in these analyses (those of
\citet{Kneiske:2004:EBL} and \citet{Primack:2001:EBL}) so this is one
of the factors contributing to the different results. Additionally,
for these previous redshift constraints using gamma-ray data, it was
necessary to make assumptions about the hardness of the intrinsic
spectrum in the HE regime. The \Fermi measurement of the HE spectrum
from PG\,1553+113 allows us to reduce the number of assumptions used
in the constraining the redshift. Combining the \Fermi measurement
with the measured VHE spectra affords us complete coverage of the
PG\,1553+113 SED from 200\,MeV to 1.1\,TeV.

Its potentially large distance together with its consistently
detectable gamma-ray flux make PG\,1553+113 an excellent candidate
with which to search for EBL cascading effects first proposed by
\citet{Protheroe:1986:EBLCascades} and, subsequently explored by many
authors (e.g., \citealp{Protheroe:1993:EBLCascades};
\citealp{Aharonian:1994:EBLCascades};
\citealp{Biller:1995:EBLCascades}; \citealp{DAvezac:2007:EBLCascades};
\citealp{Elyiv:2009:EBLCascades}). If present, pile-up from such
cascades is predicted to occur below 100\,GeV. Some authors predict
that cascades produce a characteristic spectral bump, which could be
detectable in the \Fermi energy regime if the strength of the
extragalactic magnetic field in the direction of the source being
studied is less than $B$\,=\,10$^{-6}$\,nG
\citep{DAvezac:2007:EBLCascades}. \citet{Elyiv:2009:EBLCascades}
predict that, for magnetic field strengths of
$B$\,$\leq$\,10$^{-7}$\,nG, extended emission due to the cascading of
the source photons should be detectable around extragalactic gamma-ray
sources by \Fermic.

The possibility that PG\,1553+113 could be a particularly distant TeV
source provides further impetus for IR/optical/UV astronomers to
revisit the redshift measurement. A direct measurement would be very
welcome, and would ultimately settle the question of the accuracy of
estimates based on the EBL. The \FermiLAT is continuing to accumulate
data on PG\,1553+113. Subsequent studies of these data will,
therefore, allow us to measure its spectral shape with greater
sensitivity at the highest energies accessible to \Fermic.

\section{Acknowledgments}

The \Fermi LAT Collaboration acknowledges the generous support of a
number of agencies and institutes that have supported the \Fermi LAT
Collaboration. These include the National Aeronautics and Space
Administration and the Department of Energy in the United States, the
Commissariat \`a l'Energie Atomique and the Centre National de la
Recherche Scientifique / Institut National de Physique Nucl\'eaire et
de Physique des Particules in France, the Agenzia Spaziale Italiana
and the Istituto Nazionale di Fisica Nucleare in Italy, the Ministry
of Education, Culture, Sports, Science and Technology (MEXT), High
Energy Accelerator Research Organization (KEK) and Japan Aerospace
Exploration Agency (JAXA) in Japan, and the K.~A. Wallenberg
Foundation, the Swedish Research Council and the Swedish National
Space Board in Sweden.

Additional support for science analysis during the operations phase
from the following agencies is also gratefully acknowledged: the
Istituto Nazionale di Astrofisica in Italy and the K.~A. Wallenberg
Foundation in Sweden.

This research has made use of the NASA/IPAC Extragalactic Database
(NED) which is operated by the Jet Propulsion Laboratory, California
Institute of Technology, under contract with the National Aeronautics
and Space Administration. This research has made use of the SIMBAD
database, operated at CDS, Strasbourg, France.

Thanks to Wystan Benbow, Luigi Costamante, Daniela Dorner, Andrea
Tramacere and Robert Wagner for supplying archival data-points.

{\it Facilities:} \facility{Fermi LAT}

\bibliography{myPapers}

\end{document}